\documentclass[12pt]{article}
\usepackage{color,cite,amsmath,amssymb,stmaryrd,graphicx}

\long\def\note#1{\setbox0=\hbox{\large #1}
\ifdim\wd0<0.85\textwidth
\begin{center}\framebox
{\color{blue}\quad\large #1\quad}\end{center}
\else
\begin{center}\framebox{\parbox{0.9\textwidth}
{\color{blue}\centering\large #1}}\end{center}
\fi
}

\title{\bf Number of
fermion generations from a novel Grand Unified model}
\author{\normalsize\bf 
Pritibhajan Byakti$^a$, 
David Emmanuel-Costa$^b$, Arindam Mazumdar$^a$, \\
 \normalsize\bf Palash B Pal$^a$ \\ 
 \normalsize $^{a)}$ Saha Institute of Nuclear Physics,
  Kolkata 700064, India \\ 
 \normalsize $^{b)}$ Departamento de
  F\'{\i}sica and Centro de F\'{\i}sica Te\'orica de Part\'{\i}culas
  (CFTP)\\ \normalsize Instituto Superior T\'ecnico (IST), Universidade
 de Lisboa\\ \normalsize Av. Rovisco Pais, 1049-001 Lisboa, Portugal} 
\date{}

\hsize=\textwidth
\def\Eqn#1{Eq.\ (\ref{#1})}
\def\Eqs#1#2{Eqs.\ (\ref{#1}) and (\ref{#2})}
\def\3Eqs#1#2#3{Eqs.\ (\ref{#1}), (\ref{#2}) and (\ref{#3})}

\def\sec#1{\S\,\ref{#1}}

\def\gr331{{$\rm SU(3)_c \times SU(3)_L \times U(1)_X$}}
\def\rep#1{\ensuremath{\mathbf{#1}}}
\def\Rep331#1{\ensuremath{\Big[#1 \Big]}}
\def\SMrep#1{\ensuremath{\Big\{#1 \Big\}}}
\def\form#1{\ensuremath{\llbracket \, #1 \, \rrbracket}}

\begin{document}

\maketitle

\begin{abstract}

Electroweak interactions based on a gauge group $\rm SU(3)_L \times
U(1)_X$, coupled to the QCD gauge group $\rm SU(3)_c$, can predict the
number of generations to be multiples of three.  We first try to unify
these models within $\rm SU(N)$ groups, using antisymmetric tensor
representations only.  After examining why these attempts fail, we
continue to search for an $\rm SU(N)$ GUT that can explain the number
of fermion generations.  We show that such a model can be found for
$N=9$, with fermions in antisymmetric rank-1 and rank-3
representations only, and examine the constraints on various masses in
the model coming from the requirement of unification.

\end{abstract}

\section{Introduction}
In the standard model, the number of fermion generations appears
as an arbitrary parameter, meaning that a mathematically consistent
theory can be built up using any number of fermion generations.  The
same is true for many extensions of the standard model, including
grand unification models based on the gauge groups SU(5) and SO(10).
An interesting question related to extension of the standard model is
whether the number of fermion generations can in any way be explained
through the internal consistency of the model.  In the literature,
there is some discussion of grand unification models based on big
orthogonal groups like SO(18), where one spinor multiplet contains all
known fermion fields of all generations, and much more
\cite{Wilczek:1981iz}.  It was shown that \cite{Chang:1985jd,
  Hubsch:1985zn}, with suitable symmetry breaking scheme, only three
generations can remain light, whereas others obtain masses at much
above the electroweak scale.

In a quite different line of development, it was shown that if one
extends the electroweak group to $\rm SU(3) \times U(1)$ and tries to
accommodate the standard fermions into multiplets of this gauge group
which must include some new fermions, cancellation of gauge anomalies
can restrict the number of generations and one can obtain consistent
models with the number of generations equal to three or any multiple
of it \cite{Singer:1980sw, Pisano:1991ee, Frampton:1992wt,
  Valle:1983dk, Foot:1994ym}.

These models will be described briefly in \sec{s:331}.  Then, in
\sec{s:suN} and \sec{s:N>6}, we try to see whether these models can be
embedded into a simple SU(N) group.  We conclude that, if one uses
only completely antisymmetric tensor representations, such an
embedding cannot be found.  Then, in \sec{s:other}, we start looking
for general conditions that will specify the number of fermion
generations for arbitrary SU(N) groups.  In \sec{s:su9}, we analyze
one simple model, based on the group SU(9), that gives three
generations.  The renormalization group analysis of various scales in
this models has been performed in \sec{s:gut}.  We end with some
concluding remarks in \sec{s:conclu}.

\section{The 3-3-1 models}\label{s:331}
The 3-3-1 models are based on the gauge group \gr331.  The first
factor in the gauge group is the group of QCD, whereas the other two
factors pertain to electroweak interactions.  There are two versions
of such models, and we discuss them one by one.

In one version, proposed by Pisano and Pleitez \cite{Pisano:1991ee}
and by Frampton \cite{Frampton:1992wt} (to be denoted as the PPF
model), the left-chiral fermions and antifermions belong to the
following representations of the gauge group:
\begin{subequations}
\label{PPFrep}
\begin{eqnarray}
f_a & = \left( \begin{array}{c} \hat\ell \\ \nu_\ell\\ \ell
\end{array} \right)_a & \sim \Rep331{1,3,0} 
\label{PPFrep:f} \\
Q_1 & = \left( \begin{array}{c} T_1 \\ u_1 \\ d_1
\end{array} \right) & \sim \Rep331{3,3,{2\over3}} \\
Q_i & =  \left( \begin{array}{c} B \\ d \\ u 
\end{array} \right)_i & \sim \Rep331{3,3^*,-{1\over3}} \\
\hat u_a && \sim \Rep331{3^*,1,-{2\over3}}_a \\
\hat d_a && \sim \Rep331{3^*,1,{1\over3}}_a \\
\hat T_1 && \sim \Rep331{3^*,1,-{5\over3}} \\
\hat B_i && \sim \Rep331{3^*,1,{4\over3}}_i \,.
\end{eqnarray}
\end{subequations}
Note that there are two kinds of generation indices: $a$ goes from 1
to 3, whereas $i$ takes only the values 2 and 3.  An antifermion has
been denoted by a hat.  We emphasize that all representations given
above pertain to the left-chiral components only.  The representation
of a right-chiral fermion would be the complex conjugate of that of
the left-chiral antifermion, and vice versa.  Note that there are
extra quark fields, i.e., fields which are triplets of $\rm SU(3)_c$,
but there is no neutrino field that is sterile under the standard
model.  Different generations of fermions are not copies of one
another.  Gauge anomalies cancel between the three generations
\cite{Pisano:1991ee, Frampton:1992wt}, but not within a single
generation.  Thus, the consistency of the model requires three
generations of fermions.  Of course, this pattern of three generations
can be repeated, obtained number of generations to be multiples of 3.

The other version of 3-3-1 models was proposed by Singer, Valle and
Schechter \cite{Singer:1980sw} (and hence will be referred to as the
SVS model) and was examined by other authors later \cite{Valle:1983dk,
  Foot:1994ym}.  In this version, there are sterile neutrinos, the
left-chiral component of which has been denoted by $\hat\nu$ in the
list below:
\begin{subequations}
\label{FLTrep}
\begin{eqnarray}
f_a & = \left( \begin{array}{c} \nu_\ell \\  \ell \\ \widehat \nu_\ell
\end{array} \right)_a & \sim \Rep331{1,3,-{1\over3}} \\
\hat\ell_a && \sim \Rep331{1,1,1} 
\label{FLTrep:l^}\\
Q_1 & = \left( \begin{array}{c} u_1 \\ d_1 \\ u'_1
\end{array} \right) & \sim \Rep331{3,3,{1\over3}} \\
Q_i & =  \left( \begin{array}{c} d \\ u \\ d'
\end{array} \right)_i & \sim \Rep331{3,3^*,0} \\
\hat u_a, \hat u'_1 && \sim \Rep331{3^*,1,-{2\over3}} \\
\hat d_a, \hat d'_i && \sim \Rep331{3^*,1,{1\over3}} \,.
\end{eqnarray}
\end{subequations}
The notation for the generation indices is as before.  The primed
fields are extra quark fields which are not present in the standard
model.  Like the previous model, gauge anomalies cancel between
various generations.

The difference between the two models may be summarized in the
following way.  The standard model gauge group is not a maximal
subgroup of the group \gr331.  In particular, the electroweak $\rm
SU(3)_L$ has two neutral generators, and some combination of these
two, along with the generator of $\rm U(1)_X$, form $Y$ and $I_{3L}$,
the two neutral generators of the standard electroweak model.  Using
the standard normalization of the SU(3) generators,
\begin{eqnarray}
\mathop{\rm tr} (T_A T_B) = \frac12 \delta_{AB} \,,
\label{eq:gutnorm}
\end{eqnarray}
these combinations are given by
\begin{subequations}
\label{PPF2SM}
\begin{eqnarray}
I_{3L} &=& - \frac12 T_3 + \frac{\surd3}2 T_8 \,, \\*
Y &=& \frac32 T_3 + \frac{\surd3}2 T_8 + X 
\end{eqnarray}
\end{subequations}
in the first model, whereas in the second one, they are given by
\begin{subequations}
\label{FLT2SM}
\begin{eqnarray}
I_{3L} &=& T_3 \,, \\*
Y &=& - \frac1{\surd3} T_8 + X \,.
\end{eqnarray}
\end{subequations}
The SU(3) generators are given by
\begin{eqnarray}
T_i = \begin{cases} \frac12 \lambda_i & \mbox{for the fundamental
  representation}, \\ 
-\frac12 \lambda^*_i & \mbox{for the anti-fundamental representation},
\end{cases}
\end{eqnarray}
%
where the $\lambda$'s are the well-known
Gell-Mann matrices.

\section{Seeking embeddings into SU(N)}\label{s:suN}
Since the group \gr331 is of rank 5, the smallest unitary group that
contains it as a subgroup is SU(6).  Therefore, in this section, we
analyze whether the models discussed in \sec{s:331} can be embedded
into an SU(6) grand unified model.

For the sake of convenience, let us announce here the notation that
will be used for denoting representations of various groups.  For the
grand unified group, the notations will be denoted by boldface.  The
\gr331 representations will be denoted by three numbers in
parenthesis, e.g., \Rep331{a,b,c}, as has already been done in
\Eqs{PPFrep}{FLTrep}.  And the representation of the standard model
group, when required, will be denoted by square brackets, e.g.,
$\SMrep{a,b,c}$.  Thus, with this notation, the fundamental
representation of SU(6) has the decomposition
\begin{eqnarray}
\rep 6 &\to& \Rep331{3,1,-\frac13} + \Rep331{1,3,\frac13} \nonumber\\* 
&\to& \SMrep{3,1,-\frac13} + \SMrep{1,2,\frac12} + \SMrep{1,1,0} \,.
\label{6ofsu6}
\end{eqnarray}

It is now easy to see that neither PPF nor SVS model can be embedded
into SU(6).  For the PPF model \cite{Pisano:1991ee, Frampton:1992wt},
note the representation of leptons given in \Eqn{PPFrep:f}.
Certainly, it is not contained in the fundamental representation of
SU(6) or its complex conjugate, $\rep 6^*$.  Higher representations
can be obtained by taking Kronecker products of the $\rep 6$ and $\rep
6^*$ representations, and will be of the generic form
$(\rep6)^m(\rep6^*)^n$.  Denoting the two different \gr331
representations that appear in \Eqn{6ofsu6} by $A$ and $B$, we can
write 
\begin{eqnarray}
(\rep6)^m(\rep6^*)^n &\to& \sum_{m',n'} {m \choose m'} {n \choose n'}
(A)^{m-m'} (B)^{m'} (A^*)^{n-n'} (B^*)^{n'} \,.
\end{eqnarray}
Take any term in the sum.  Contributions to the $\rm U(1)_X$
quantum number come from all four factors, and is given by
\begin{eqnarray}
X = \frac13 ( -m + 2m' + n - 2n') \,.
\end{eqnarray}
Thus, for the lepton representation such as in \Eqn{PPFrep:f}, we need 
\begin{eqnarray}
m-n = 2(m'-n') \,.
\end{eqnarray}
in order to obtain $X=0$.  Only the $B$ and $B^*$ contributes to
non-trivial $\rm SU(3)_L$ representations.  In order to obtain a
triplet, we need
\begin{eqnarray}
m' - n' = 1 \mod 3 \,,
\end{eqnarray}
considering the triality of the representations.  Moreover, the lepton
must be a color singlet, which means that we should have
\begin{eqnarray}
(m-m') - (n-n') = 0 \mod 3 \,.
\end{eqnarray}
These three conditions cannot be satisfied with integers, and hence it
is impossible to obtain a $(1,3,0)$ representation of \gr331 in any
representation of SU(6).

For the SVS model, the same kind of analysis can be performed
keeping an eye towards the antilepton in \Eqn{FLTrep:l^}.  In order to
produce a singlet of both SU(3) factors, one needs Kronecker product
of equal number of $6$ and $6^*$.  However, such products will give
$X=0$.

We have thus proved that neither the PPF, nor the SVS model can be
embedded into SU(6).  This result should not be understood to imply
that an SU(6) grand unified model is impossible.  We can take a
different embedding of the standard model generators $I_{3L}$ and $Y$,
viz.,
\begin{subequations}
\begin{eqnarray}
I_{3L} &=& T_3 \,, \\*
Y &=& \frac1{\surd3} \lambda_8 +  X \,.
\end{eqnarray}
\end{subequations}
The SM reduction of the $6^*$ representation of SU(6) can be easily
read from \Eqn{6ofsu6}:
\begin{eqnarray}
\rep 6^* \to \underbrace{\SMrep {3^*,1,\frac13}}_{\hat d}  + 
\underbrace{ \SMrep{ 1,2,-\frac12}}_{L} + \SMrep{ 1,1,0 } \,.
\label{6*su6}
\end{eqnarray}
Also, the antisymmetric rank-2 representation has the following
decomposition under the SM gauge group:
\begin{eqnarray}
\rep {15} &\to& \Rep331{3^*,1, -\frac23} +
\Rep331{1,3^*,\frac23} + \Rep331{3,3,0} \nonumber\\
&\to& 
\underbrace{ \SMrep{3^*,1,-\frac23}}_{\hat u} + \SMrep{1,2,\frac12} +
\underbrace{ \SMrep{1,1,1}\vphantom{1\over6}}_{\hat\ell} + 
\underbrace{ \SMrep{3,2,\frac16}}_{Q} +
\SMrep{3,1,-\frac13} \,.
\label{15su6}
\end{eqnarray}
In both \Eqs{6*su6}{15su6}, we have marked the known fermions which
correspond to the SM representations.  We observe that all known
fermions of a single generation belong to these two representations.
This is therefore like the minimal SU(5) grand unified model where the
corresponding representations contain all known fermion fields of a
single generation and nothing more.  In this case, there are some
extra fermions to complete the SU(6) representations.

There is one big difference between the SU(5) and this SU(6) model.
This can be seen from the anomaly coefficients of different
representations.  For the completely antisymmetric tensorial
representations of SU(N), the anomaly coefficients are as follows
\cite{Banks:1976yg}:
\begin{eqnarray}
\begin{array}{c|ccc}
\hline
\mbox{Representation} & \form 1 & \form 2 & \form 3 \\
\hline  
\mbox{anomaly coefficient} & 1 & N-4 & \frac12 (N-3)(N-6) \\
\hline
\end{array}
\label{anom}
\end{eqnarray}
Here, the completely antisymmetric representation of rank $n$ has been
denoted by \form n.  The anomaly coefficient of any representation and
its complex conjugate will be negative of each other.  We see that for
the SU(5) group, anomalies cancel between the $\rep5^*$ and $\rep{10}$
representations.  For the SU(6) gauge group, the $\rep 6^*$ and the
$\rep{15}$ representations have anomaly coefficients $-1$ and 2
respectively.  Thus, along with 3 copies of the $\rep{15}$
representation that is necessary in order to have three quark
doublets, we need 6 copies of the $\rep 6^*$ representation in order
to cancel anomalies.  There will thus be 3 extra copies of the lepton
doublet $L$.  But these will form gauge invariant masses with the
three copies of the $\SMrep{1,2,\frac12}$ representation that appears
in the three $\rep{15}$-plets, and these masses can be much heavier
than the electroweak scale.  Similarly, among the six copies of the
$\hat d$ representation that appear in the $\rep 6^*$ multiplets,
three will form gauge invariant masses with the $\SMrep{3,1,-\frac13}$
representations present in the $\rep{15}$-plets, leaving only three
$\hat d$'s for the electroweak scale.  The sterile neutrino fields,
i.e., the $\SMrep{1,1,0}$ representations shown in \Eqn{6*su6}, can
also have mass terms that are invariant under the SM gauge group.  If
all these masses are large, the only fields that are left over at the
electroweak scale are no different from the fields that are obtained
in three generations of $\rep5^*$ and $\rep{10}$ multiplets of SU(5).
Moreover, the model has no explanation for the number of generations.
Any number of $\rep{15}$-plets, along with twice the number of
$\rep6^*$-plets, would be anomaly-free.  Hence this model is not
interesting for our discussion.

\section{Generalities about SU(N) models with $N>6$}\label{s:N>6}
We now consider models based on the gauge groups SU(N), with $N>6$.
Henceforth we will use completely antisymmetric tensor representations
only.  Such a representation of rank $n$ will be denoted by \form n,
as was done in \Eqn{anom}.  The fundamental representation will be
thus denoted by \form 1 in this notation, whereas its complex
conjugate will be \form{N-1}.  We will show that, with antisymmetric
representations only, neither PPF nor SVS model can be embedded into
an SU(N) grand unified group.

The crucial aspect of both PPF and SVS models is that, among the
left-chiral fields, the quarks, i.e., color triplets, appear in
triplet or antitriplet of $\rm SU(3)_L$, whereas the antiquarks, i.e.,
color antitriplets, are all singlets of $\rm SU(3)_L$.  In particular,
then, there is no multiplet that transforms like \Rep331{3^*,3,\star}
or \Rep331{3^*,3^*,\star}.  On the other hand, among the quark fields,
some should be in \Rep331{3,3,\star} and some in \Rep331{3,3^*,\star}
representations, thereby ensuring anomaly cancellation among different
generations.  These features are not available in the decomposition of
any \form m representation of an SU(N) group, as we show now.

Since \gr331 is not a maximal subgroup of the groups under
consideration, it should be possible to embed the \gr331 group into
the SU(N) grand unified group in more than one ways.  First, we assume
that the decomposition of the fundamental representation of SU(N) into
\gr331 is given by
\begin{eqnarray}
\form 1 \equiv \rep N \to \Rep331{3,1,\star} + \Rep331{1,3,\star} +
\sum_{k=7}^N  \Rep331{1,1,\star} \,,
\label{NofSUN}
\end{eqnarray}
where the $\rm U(1)_X$ charges have been left unspecified, denoted by
the star symbol.  There are various ways of assigning the $\rm U(1)_X$
quantum numbers, and the specifics are irrelevant for our discussion.
The representation \form m can contain a \Rep331{3^*,3,\star}
submultiplet if, among the $m$ tensor indices, 2 come from the color
part and 1 from the $\rm SU(3)_L$ part, and the remaining $m-3$
indices should belong to the $\rm U(1)_X$ subgroup.  However, the
count of \Rep331{3,3^*,\star} is exactly the same.  Thus, we obtain
equal numbers of \Rep331{3^*,3,\star} and \Rep331{3,3^*,\star}
submultiplets.  The only way to get rid of the \Rep331{3^*,3,\star}
submultiplets is for them to form \gr331 invariant masses with
\Rep331{3,3^*,\star} submultiplets and become superheavy.  But then
there is no remaining \Rep331{3,3^*,\star} submultiplet to contribute
to the fermion content of PPF or SVS models.  Hence the impasse.

The situation is no different if we assume that the decomposition of
the fundamental follows the rule
\begin{eqnarray}
\form 1 \equiv \rep N \to \Rep331{3,1,\star} + \Rep331{1,3^*,\star} +
\sum_{k=7}^N  \Rep331{1,1,\star} \,.
\label{N2ofSUN}
\end{eqnarray}
In this case, following the same argument, we would conclude that the
number of \Rep331{3,3,\star} and \Rep331{3^*,3^*,\star} are equal.
Now, the \Rep331{3^*,3^*,\star} submultiplets can be got rid of by
forming superheavy masses with \Rep331{3,3,\star} submultiplets.  But
then there will be no \Rep331{3,3,\star} left at the \gr331 level,
which is unacceptable for both PPF and SVS models.

\section{Other embeddings into SU(N)}\label{s:other}
At this point, we ignore the intermediate symmetry \gr331 and try to
see whether it is possible to embed the standard model fermions into
an SU(N) grand unified group that would provide an explanation for the
number of fermion generations.  To this end, we first list the
decomposition of the lowest rank antisymmetric tensor representations
of SU(N) into the SM gauge group.  The fundamental representation
decomposes as follows:
\begin{eqnarray}
\label{breaking}
\form 1 &=& \SMrep{3,1,-\frac13} + \SMrep{1,2,\frac12} +
(N-5) \cdot \SMrep{1,1,0} \,,
\label{eq:t1}
\end{eqnarray}
where the number within parentheses indicate the number of copies of
the SM gauge singlet in the last term.  Similarly, we obtain
\begin{eqnarray}
\form 2 &=& \SMrep{3^*,1,-\frac23} + \SMrep{3,2,\frac16} + (N-5) \cdot
\SMrep{3,1,-\frac13} \nonumber\\* 
&& \null + \SMrep{1,1,1} + (N-5) \cdot
\SMrep{1,2,\frac12} + {N-5 \choose 2} \cdot \SMrep{1,1,0} \,, \nonumber\\ 
\form 3 &=& \SMrep{1,1,-1} + \SMrep{3^*,2,-\frac16} + (N-5) \cdot
\SMrep{3^*,1,-\frac23} \nonumber\\* 
&& \null + (N-5) \cdot \SMrep{3,2,\frac16} +
\SMrep{3,1,\frac23} + {N-5 \choose 2} \SMrep{3,1,-\frac13} \nonumber\\* 
&& \null + (N-5) \cdot \SMrep{1,1,1} + {N-5 \choose 2} \SMrep{1,2,
  \frac12} + {N-5 \choose 3} \SMrep{1,1,0} \,.
\label{eq:t2t3}
\end{eqnarray}
\begin{table}
\caption{Number of different SM multiplets occurring in completely
  antisymmetric representations of SU(N).}\label{t:SMreps}

$$
\begin{array}{c|ccc}
\hline
& \multicolumn{3}{c}{\mbox{Representation}} \\ 
\cline{2-4}
\mbox{SM multiplet} & \form 1 & \form 2 & \form 3 \\
\hline  
\SMrep{1,2,-\frac12} & -1 & -(N-5) & -{N-5 \choose 2} \\
\SMrep{3,2,\frac16} & 0 & 1 & N-6 \\
\SMrep{3^*,1,-\frac23} & 0 & 1 & N-6 \\
\SMrep{3^*,1,\frac13} & -1 & -(N-5) &  -{N-5 \choose 2} \\
\SMrep{1,1,1} & 0 & 1 & N-6 \\ 
\hline
\end{array}
$$
\end{table}
Suppose now we consider a model with $n_1$ copies of \form 1
representation, $n_2$ copies of \form 2, and $n_3$ copies of \form 3.
In counting the ``copies'', we will denote a complex conjugate
representation by a negative number.  We count the number of different
SM multiplets in the antisymmetric representations of SU(N) and
summarize the result in Table\,\ref{t:SMreps}.  We see that, in order
to obtain $n_g$ generations of fermions, we need 
\begin{subequations}
\label{conditions}
\begin{eqnarray}
n_2 + (N-6) \,n_3 = n_g 
\label{gen1}
\end{eqnarray}
in order to ensure the correct number of quark doublets, as well as
$\hat u$ and $\hat\ell$.  In addition, we need 
\begin{eqnarray}
n_1 + (N-5) n_2 + {N-5 \choose 2} \,n_3 = - n_g 
\label{gen2}
\end{eqnarray}
so that the correct number of lepton doublets and $\hat d$ are
obtained.  Anomaly cancellation between the representations will be
ensured if we have
\begin{eqnarray}
n_1 + (N-4) n_2 + \frac12 (N-3)(N-6) \,n_3 = 0 \,,
\label{anomcancel}
\end{eqnarray}
\end{subequations}
making use of the anomaly co-efficients given in \Eqn{anom}.  However,
only two of the three relations in \Eqn{conditions} are independent.
We find it simple to work with \Eqs{gen1}{anomcancel}.  The general
solution of these equations is given by
\begin{eqnarray}
n_1 &=& -(N-4) n_g + \frac12 (N-5)(N-6) \,n_3 \,, \nonumber\\*
n_2 &=& n_g - (N-6) \,n_3 \,.
\label{nsolu}
\end{eqnarray}

If we take $n_3=0$, we obtain $n_g=n_2$, and hence no explanation of
generations.  This is what was done for the SU(6) grand unified model
discussed in Sec.\,\ref{s:suN}, a model which was found uninteresting
precisely because it could not predict the number of generations.
However, we can consider other kinds of solutions of \Eqn{nsolu}.  For
example, if we take $n_2=0$, we obtain 
\begin{eqnarray}
n_g = (N-6) \,n_3 \,.
\end{eqnarray}
In this case, for the grand unified group SU(9), we find that the
number of generations must be 3 or its multiple.

It should be noted that it is just as easy to obtain solutions of
\Eqn{nsolu} with $n_1$, $n_2$ and $n_3$ all non-zero.  One such model
with an SU(8) gauge group was the subject matter of
Ref.\cite{Martinez:2011iq}, where the authors took $n_1=-9$, $n_2=1$,
$n_3=1$ and obtained three generations of fermions.  From our
analysis, it seems that they could have obtained any other number of
generations by adjusting the number of copies of various
representations.  For example, $n_1=-13$, $n_2=2$, $n_3=1$ would give
four generations.  However, the merit of the $n_2=0$ solutions is that
the number of generations cannot be arbitrary: it must be a multiple
of 3.

\section{Anatomy of an SU(9) model}\label{s:su9}
As seen from our earlier analysis, an SU(9) model, in the absence of 
rank-2 antisymmetric multiplets, automatically gives three
fermion generations, provided we take
\begin{eqnarray}
n_3 = 1 \,, \qquad n_1 = -9 \,,
\label{-9+1}
\end{eqnarray}
which means that there should be 9 copies of the anti-fundamental
representation, and one copy of the rank-3 antisymmetric multiplet.

To see in more detail how different fermion representations of the
standard model are obtained from these representations of the grand
unified group SU(9), we first discuss the decomposition of these
multiplets under the group \gr331.  The decomposition of the
fundamental of SU(9) can be taken to be as given in \Eqn{NofSUN}.
We can choose the $\rm U(1)_X$ quantum number in a way that it
vanishes for all singlets of $\rm SU(3)_c \times SU(3)_L$.  Then,
choosing the normalization of the $\rm U(1)_X$ arbitrarily, we can
write 
\begin{eqnarray}
\form {\bar 1} \equiv \rep {\bar 9} \to \Rep331{3^*,1,\frac13} +
\Rep331{1,3^*,-\frac13} + 
3 \cdot \Rep331{1,1,0} \,.
\end{eqnarray}
This gives, for the rank-3 representation of SU(9), the following
decomposition into \gr331 multiplets: 
\begin{eqnarray}
\form 3 \equiv \rep {84} &=& \Rep331{1,1,0} + \Rep331{1,1,1} +
\Rep331{1,1,-1} \nonumber\\*
&& + 3 \cdot \Rep331{3,3,0} + \Rep331{3^*,3,-\frac13} +
\Rep331{3,3^*,\frac13} \nonumber\\* 
&& + 3 \cdot \Rep331{3^*,1,-\frac23} + 3 \cdot \Rep331{1,3^*,\frac23} 
+ 3 \cdot \Rep331{1,3,\frac13} + 3 \cdot \Rep331{3,1,-\frac13} . 
\end{eqnarray}
Looking at these decompositions, we find that the multiplets specified
by \Eqn{-9+1} contain the following vector-like combinations of \gr331
submultiplets: 
\begin{eqnarray}
\begin{array}{r@{\hspace{7mm}}l}
28 & \Rep331{1,1,0} \\[2mm]
3 & \Rep331{3^*,1,\frac13} + \Rep331{3,1,-\frac13} \\[2mm] 
3 & \Rep331{1,3^*,-\frac13} + \Rep331{1,3,\frac13} \\[2mm]
1 & \Rep331{3^*,3,-\frac13} + \Rep331{3,3^*,\frac13}  \\[2mm]
1 & \Rep331{1,1,1} + \Rep331{1,1,-1} \,. \\ 
\end{array}
\label{331vec}
\end{eqnarray}
Such things can have \gr331 invariant mass terms at the level where
the said symmetry is intact, and do not affect the SM reduction of the
model.  These singlets and vector-like particles do not contribute to
the \gr331 anomalies.

In addition, we find that the decomposition consists of several chiral
multiplets of \gr331.  Under the group $\rm SU(3)_L$, these multiplets
transform either like triplets or antitriplets, or like singlets.  We
want the triplets and antitriplets to contain the doublets of the
standard electroweak gauge group.  For that, we first have to discuss
how the SM is embedded into \gr331.  Since the members of an $\rm
SU(2)_L$ doublet have the same value of hypercharge $Y$, we need to
ensure that the diagonal $Y$ generator has two equal entries in the
$3\times3$ representation of $\rm SU(3)_L$.  We notice that the PPF
model was constructed in such a way that the second and the third
elements of a triplet end up with the same value of $Y$ through
\Eqn{PPF2SM}, whereas in the SVS model, the first two elements are
equal.  In order to obtain something different, we can now try a
solution in which the first and the third elements of $Y$ should have
the same diagonal elements.  This solution is given by
\begin{subequations}
\label{our2SM}
\begin{eqnarray}
I_{3L} &=& \frac12 T_3 + \frac{\surd3}2 T_8 \,, \\*
Y &=& \frac12 T_3 - \frac1{2\surd3} T_8 + X \,.
\end{eqnarray}
\end{subequations}
With these assignments, we now present the chiral multiplets of \gr331
that are present in the choice of representations in \Eqn{-9+1}, as
well as their SM decompositions:
\begin{subequations}
\label{SMsurvivors}
\begin{eqnarray}
3 \cdot \Rep331{3,3,0} &\to& 3 \cdot \SMrep{3,2,\frac16} + 3 \cdot
\SMrep{3,1,-\frac13} \,, 
\label{330->SM} \\* 
6 \cdot \Rep331{3^*,1,\frac13} &\to& 6 \cdot \SMrep{3^*,1,\frac13}
\,, 
\label{3*1d}\\* 
6 \cdot \Rep331{1,3^*,-\frac13}  &\to& 6 \cdot \SMrep{1,2,-\frac12} +
6 \cdot \SMrep{1,1,0} \,, \label{} \\* 
3 \cdot \Rep331{3^*,1,-\frac23}  &\to& 3 \cdot \SMrep{3^*,1,-\frac23}
\,, \\* 
3 \cdot \Rep331{1,3^*,\frac23} &\to& 3 \cdot \SMrep{1,2, \frac12} +
3 \cdot \SMrep{1,1,1} . 
\end{eqnarray}
\end{subequations}

We see that, at the SM level, there are three quark doublets, three
multiplets that transform like $\hat u_L$, and also three that
transform like the $\hat\ell_L$: precisely the numbers necessary for
obtaining three fermion generations in the SM.  We find six SM
multiplets that transform like the $\hat d_L$ in \Eqn{3*1d}, but that
is not a problem, because three of them can pair up with and equal
number of \SMrep{3,1,-\frac13} multiplets that appear in
\Eqn{330->SM}, and can obtain a bare mass term that is invariant under
the SM gauge group.  Similarly, the three \SMrep{1,2, \frac12}
multiplets pair up with three of the six \SMrep{1,2, -\frac12}
multiplets, leaving three lepton doublets at the SM level.  Thus,
we are left with exactly three chiral generation of SM fermions which
are necessary for building up the standard model.  In addition, there
are  vector-like combinations of the SM gauge group which come from
\Eqn{331vec} and \Eqn{SMsurvivors}.  We write these combinations as
follows: 
\begin{eqnarray}
\begin{array}{clc}
\hline
\mbox{Notation} & \mbox{Representations under} &
\mbox{Number of} \\ 
& \mbox{the SM gauge group} & \mbox{copies} \\ 
\hline\\[-1mm]
Q & \SMrep{3,2, \frac16} +\SMrep{3^*,2,- \frac16} & \ 1  \\[2mm]
U &  \SMrep{3,1, \frac23} +\SMrep{3^*,1,- \frac23} & \ 1  \\[2mm] 
D &  \SMrep{3,1,- \frac13}+\SMrep{3^*,1,- \frac16} & \ 6  \\[2mm] 
L &  \SMrep{1,2, \frac12} +\SMrep{1,2,- \frac12} & \ 6  \\[2mm]
E & \SMrep{1,1,-1}+\SMrep{1,1,1} & \ 1  \\[2mm]
S & \SMrep{1,1,0} & 40 \\[4mm]
\hline
\label{SMvec}
\end{array}
\end{eqnarray}

This model is therefore very different from the existing models based
on the gauge group \gr331, the ones outlined in \sec{s:331}.  In the
existing models, cancellation of gauge anomalies is obtained by making
one fermion generation different from the others.  In the present
model, this is not the case.  At the level of the \gr331 group, all
fermion generations have the same transformation properties.  However,
there are extra fermions that help cancel the anomalies, fermions
which become vector-like at the SM level.  Thus, even apart from the
grand unification prospects via SU(9), the multiplets presented in
\Eqn{SMsurvivors} can be taken to represent a new \gr331 model, which
can be studied in its own right.  However, if we consider it purely as
a \gr331 model, there is no explanation of the number of generations.
One can obtain any number of generations by changing the number of
copies of each multiplet, keeping the ratios intact, and obtain any
number of generation one wants.  Thus, it is fair to say that the
explanation of the number of generations comes directly from the grand
unification group SU(9), and not from its subgroup \gr331.

\section{Gauge coupling unification and the intermediate
  scales}\label{s:gut}
One addresses now the unification picture of the SU(9) model defined
by the condition given in \Eqn{-9+1}.  For the sake of simplicity, we
shall assume in our analysis that the SU(9) GUT model spontaneously
breaks, through the scheme given by \Eqn{breaking}, directly to
the SM gauge group at a unique unification scale $\Lambda$. Thus, the
unification condition for the SM gauge couplings $\alpha_{1,2,3}$ at the
scale $\Lambda$ reads as
\begin{eqnarray}
  \label{eq:unif}
  \alpha_U = k_1\,\alpha_1(\Lambda) \,=\, k_2\,\alpha_2(\Lambda) 
  \,=\, k_3\, \alpha_3(\Lambda) \,.
\end{eqnarray}
The normalization constants $k_i$ are defined by
\begin{eqnarray}
k_i\,\equiv\, \frac{\mathop{\rm tr} T^2_i}{\mathop{\rm tr} T^2}\,,
\end{eqnarray}
where $T$ and $T_i$ are the same unbroken generator properly
normalized to the GUT group and its subgroup $G_i$, respectively, for
a given representation. Taking into account the decomposition of the
fundamental representation given in \Eqn{breaking} and the GUT
normalization in \Eqn{eq:gutnorm}, one easily derives that the
normalization constants $k_i$ take the canonical values
\begin{eqnarray}
k_1=\frac53\,,\quad  k_2=k_3=1\,,
\end{eqnarray}
like in the SU(5) GUT models. The relation given in \Eqn{eq:unif} is
only valid at the unification scale and one has to relate these gauge
coupling with measurable quantities at electroweak scale. 

The evolution of the gauge couplings in the one-loop approximation is
ruled by the solutions of the Renormalization Group Equations (RGE's),
which depend on the masses of the particles in the model.  We take the
SM particles and $n_H$ Higgs doublets at the electroweak scale.  In
addition, the vector-like combinations given in \Eqn{SMvec} can also
have masses between $M_Z$ and the unification scale $M_U$.  We will
assume, for the sake of simplicity, that all extra vector-like
fermions with the same quantum number share the same mass scale, and
denote these scales collectively as $M_I$, where $I$ can take the
`values' $Q,\,U,\,D,\,L,\,E$ as shown in \Eqn{SMvec}.  The solutions
of the RGE's can then be written as
\begin{subequations}
\begin{align}
\label{eq:RGEsol1}
\alpha^{-1}_1(\mu)&=\alpha^{-1}_1(M_Z)-\frac{b_1}{2\pi}\ln\left(\frac{\mu}{M_Z}\right)
-\sum_I\frac{b^I_1}{2\pi}\ln\left(\frac{\mu}{M_I}\right)
\,,\\
\label{eq:RGEsol2}
\alpha^{-1}_2(\mu)&=\alpha^{-1}_2(M_Z)-\frac{b_2}{2\pi}\ln\left(\frac{\mu}{M_Z}\right)
-\sum_I\frac{b^I_2}{2\pi}\ln\left(\frac{\mu}{M_I}\right)
\,,\\
\label{eq:RGEsol3}
\alpha^{-1}_3(\mu)&=\alpha^{-1}_3(M_Z)-\frac{b_3}{2\pi}\ln\left(\frac{\mu}{M_Z}\right)
-\sum_I\frac{b^I_3}{2\pi}\ln\left(\frac{\mu}{M_I}\right)
\,,
\end{align}
\end{subequations}
where $b_i$'s are the one-loop beta coefficients that take into account
the quantum numbers of the SM fermions, the gauge bosons and $n_H$
Higgs doublets:
\begin{eqnarray}
b_1=\frac{20}{9}\, n_g+\frac{n_H}{6}\,,\quad
b_2=\frac{4}{3}\, n_g+\frac{n_H}{6}-\frac{22}{3}\,,\quad
b_3=\frac{4}{3}\, n_g-11\,.
\end{eqnarray}
The $b^I_i$'s are the one-loop beta coefficients for the
intermediate extra vector-like fermions.  The values of these
$b^I_i$'s are given in table~\ref{t:betavec}.

\begin{table}
  \caption{Beta coefficients for the extra vector-like
    fermions.}\label{t:betavec}
\begin{center}
\begin{tabular}{c|ccccc}
 & $Q$ & $U$ & $D$ & $L$ & $E$\\
\hline
$b_1$ & $2/9$   & $16/9$   & $4/9$   & $2/3$ & $4/3$ \\
$b_2$ & $2$   & $0$    & $0$   & $2/3$ & $0$ \\
$b_3$ & $4/3$ & $2/3$  & $2/3$ & $0$   & $0$  \\
\end{tabular}
\end{center}
\end{table}

In order to get some insight into the unification in the one-loop
approximation, i.e.  to understand the intermediate scales which lead
to successful unification, let us define the effective beta
coefficients $B_i$~\cite{Giveon:1991zm,EmmanuelCosta:2011wp},
\begin{eqnarray}
B_i\equiv\frac{1}{k_i}\left(b_i+\sum_I b_i^I\,r_I\right)\,,
\end{eqnarray} 
where the ratio $r_{I}$ is given by
\begin{eqnarray} 
r_I =
\frac{\ln\left(\Lambda/M_I\right)}{\ln\left(\Lambda/M_Z\right)}\,.
\end{eqnarray} 
It is also convenient to introduce the differences $B_{ij}\equiv
B_i-B_j$, such that
\begin{eqnarray}
B_{ij}= B^{\text{SM}}_{ij}+\sum_I\Delta^I_{ij}r_I\,,
\end{eqnarray}
where $B^{\text{SM}}_{ij}$ corresponds to the SM particle contribution
and
\begin{eqnarray} 
\Delta^I_{ij}= \frac{b^I_i}{k_i}-\frac{b^I_j}{k_j}\,.
\end{eqnarray}
We then find that 
\begin{subequations}
\label{Btests}
\begin{eqnarray} 
\label{eq:Btest}
B\equiv\frac{B_{23}}{B_{12}} &=& 
\frac{\sin^2\theta_W-\dfrac{k_2}{k_3}\dfrac{\alpha}
{\alpha_s}}
{\dfrac{k_2}{k_1}-\left(1+\dfrac{k_2}{k_1}\right)\sin^2\theta_W}\,,
\\ 
\label{eq:Ltest} 
B_{12}\, \ln \left(\frac{\Lambda}{M_Z}\right) &=& \frac{2\pi}{\alpha}
\left[\frac{ 1}{k_1}-\left(\frac{1}{k_1}+\frac{1}{k_2}
  \right)\sin^2\theta_W\right ].
\end{eqnarray}
\end{subequations}
\begin{figure}
\begin{center}
\includegraphics{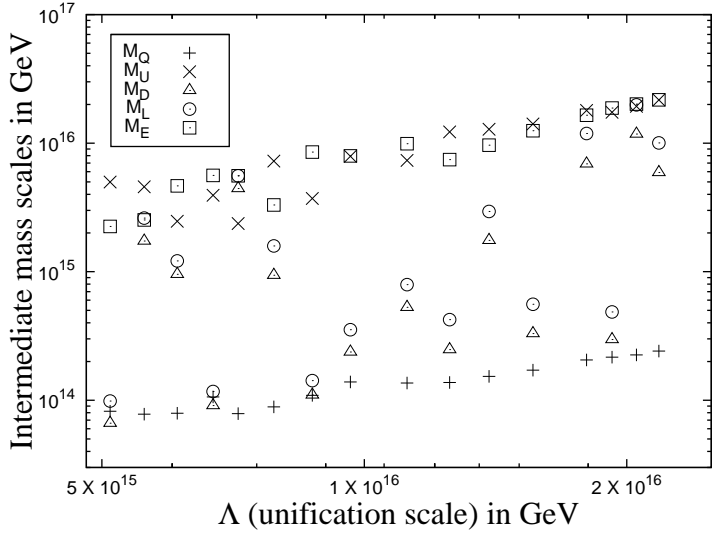}
\end{center}
\caption{\label{fig:vl} The intermediate scales of the extra
  vector-like fermions in function of the unification scale
  $\Lambda\,$.}\label{f:btest}
\end{figure}
Notice that the right-hand sides of \Eqn{Btests} depend only on
low-energy electroweak data and the group factors $k_i$. Adopting the
following experimental values at $M_Z$~\cite{Beringer:1900zz}
\begin{subequations}
\label{expvalues}
\begin{eqnarray} 
\alpha^{-1} &=& 127.916\pm0.015\,, \\
\sin^2\theta_W &=& 0.23116\pm0.00012\,, \\ 
\alpha_s &=& 0.1184\pm0.0007\,,
\end{eqnarray}
\end{subequations}
the above relations give
\begin{eqnarray}
\label{eq:Btestexp}
B &=& 0.718\pm0.003\,, \\*
B_{12}\,\ln\left(\frac{\Lambda}{M_Z}\right) &=& 185.0\pm0.2\,,
\end{eqnarray} 
in the canonical GUT models with $k_i=(5/3,1,1)$.  The coefficients
$B_{ij}$ that appear in the left-hand sides of \Eqn{Btests} strongly
depend on the particle content of the theory. For instance,
considering the SM-like particles, with $n_g$ generations, together
with $n_H$ light Higgs doublets, one has for the coefficients $B_{12}$
and $B_{23}$:
\begin{eqnarray}
\label{eq:beffSM}
B_{12}=\frac{22}{3}-\frac{n_H}{15}\,,\quad B_{23}=\frac{11}{3}+
\frac{n_H}{6}\,.
\end{eqnarray} 
In the case of SM, i.e. $n_g=3$ and $n_H=1$  one has 
\begin{eqnarray}
B\,=\,115/218\approx0.53\,,
\end{eqnarray}
that is not compatible with the calculated value in
eq.~\eqref{eq:Btestexp} and clearly, the B-test fails badly in the
SM. In table~\ref{tab:betacoeffs}, we have summarized the
contributions of the vector-like fermions to $B_{12}$ and $B_{23}$.  
\begin{table}
  \caption{\label{tab:betacoeffs} Beta coefficients for the extra
    vector-like fermions.}
\begin{center}
\begin{tabular}{c|ccccc}
 & $Q$ & $U$ & $D$ & $L$ & $E$\\
\hline
$B_{12}$ & $-28/15$ & $16/15$ & $4/15$ & $-4/15$ & $4/5$\\
$B_{23}$ & $2/3$ & $-2/3$ & $-2/3$ & $2/3$ & $0$
\end{tabular}
\end{center}
\end{table}
Using these values, we ave taken random values of the intermediate
scales of the extra vector-like fermions and found combinations of
these intermediate scales which are compatible with successful
unification.  In our numerics we have assumed only one Higgs doublet,
i.e. $n_H=1$, and we have taken a rough lower bound on the unification
scale, $M_U > 6 \times10^{15}$\,GeV, coming from the unobservability of
proton decay \cite{Beringer:1900zz} into $e^+\pi^0$.  A million set of
such random combinations were taken.  A few of the allowed ones are
presented in Fig.~\ref{f:btest}.  From the entire set of combinations
used in our numerics, we find that the allowed range for the
unification scale obtained is
\begin{eqnarray}
6\times10^{15}\,\text{GeV}\leq\Lambda\leq2.2\times10^{16}\,\text{GeV}\,,
\end{eqnarray}
which is also roughly what the limited data of Fig.~\ref{f:btest}
indicates.  For the scales of the vector-like extra fermions we have
\begin{subequations}
\begin{eqnarray}
5.5\times10^{13}\,\text{GeV}\leq
M_Q\leq2.4\times10^{14}\,\text{GeV}\,,\\[2mm] 
1.2\times10^{15}\,\text{GeV}\leq M_U
\leq2.2\times10^{16}\,\text{GeV}\,,\\[2mm] 
6.6\times10^{13}\,\text{GeV}\leq M_D
\leq1.2\times10^{16}\,\text{GeV}\,,\\[2mm] 
7.4\times10^{13}\,\text{GeV}\leq
M_L\leq2.0\times10^{16}\,\text{GeV}\,,\\[2mm] 
1.7\times10^{15}\,\text{GeV}\leq
M_E\leq2.1\times10^{16}\,\text{GeV}\,. 
\end{eqnarray}
\end{subequations}
We see that the range of mass scales of the vector-like fermions are
high, but they are indeed necessary for driving the evolution of the
gauge couplings to a perfect unification.  If one varies the number of
Higss doublet, $n_H=2,3$, the ranges of intermediate scales above do
not change significantly.

\section{Conclusions}\label{s:conclu}
To summarize, we have succeeded in building a grand unified model
based on the group SU(9).  The model uses fermions in antisymmetric
representations only and the consistency of the model demands that the
number of fermion generations is three.  As mentioned in the
introduction, earlier models based on the gauge group \gr331 and
SO(18) also had this property of predicting the number of generations.
It is interesting to note that our group SU(9) contains \gr331, and is
contained in SO(18).  However, in our model, the transformation
properties of known fermions in the \gr331 subgroup of SU(9) is not
the same as those used in earlier models based on the \gr331 group.
It is a different one, where the known fermions of all generations
transform in the same way under \gr331.  On the other hand, a
comparison with the SO(18) models also reveals an interesting
connection.  In SO(18) models, the fermion generations are contained
in the spinor representation.  The spinor representation of SO(18),
which is $\rep{256}$-dimensional, decomposes under the SU(9) subgroup
as follows:
\begin{eqnarray}
\rep{256} = \form 1 + \form 3 + \form 5 + \form 7 + \form 9 \,.
\end{eqnarray}
Our SU(9) model uses only the first two kinds of these submultiplets
to predict the number of fermion generations.

We have also shown that our model can be consistent with unification
requirements.  In this part of the analysis, we have assumed a direct
breaking of SU(9) into the SM gauge group.  A more detailed analysis,
including possibilities of intermediate symmetry breaking scales, will
be taken up in a future work.

\section*{Acknowledgments}
The work of D.E.C.~was supported by Associa\c c\~ao do Instituto
Superior T\'ecnico para a Investiga\c c\~ao e Desenvolvimento (IST-ID)
and also by FCT through the projects PEst-OE/FIS/UI0777/2011 (CFTP),
CERN/FP/123580/2011 and PTDC/FIS-NUC/0548/2012.

\bibliographystyle{unsrt}
\bibliography{gut331.bib}

\end{document}